\documentstyle[psfig,epsf,rotate]{mn}

\newcommand{\kms}{ \mbox{\,km\,s$^{-1}$}}

\newcommand{\vsini}{ V_{\rm rot} \sin~i } 
\newcommand{\Teff}{ T_{\rm eff} }
\newcommand{\Msun}{ \rm M_{\odot} }

\title[Determining the spectroscopic mass ratio in interacting binaries]
{Determining the spectroscopic mass ratio in interacting binaries:
Application to X-Ray Nova~Sco~1994}

\author[T.~Shahbaz] 
{T.~Shahbaz \\ 
Instituto de Astrof\'\i{}sica de Canarias E-38200 La Laguna, 
Tenerife, Spain }

\begin{document}

\maketitle

\begin{abstract}

We present a model for determining the mass ratio in interacting binaries
by directly fitting the observed spectrum with synthetic spectra.  We make
direct use of $\sc NextGen$ model atmospheres intensities which are the
most comprehensive and detailed models available for cool stars. We fully
take into account the varying temperature and gravity across the
secondary star's photosphere, by incorporating the synthetic spectra into
the secondary star's Roche geometry. As a result, we determine the exact
rotationally broadened spectrum of the secondary star and so eliminate the
need for a limb-darkening law, and the uncertainties associated with it.

As an example we determine the  mass ratio for the well studied soft X-ray
transient Nova~Sco~1994.    In order to obtain a more accurate determination of
the mass ratio, which does not depend  on assumptions about the rotation
profile and limb-darkening coefficients,  we use our model to compute the exact
rotationally broadened model spectrum,  which we compare directly  with the
observed intermediate resolution spectrum of Nova~Sco~1994.
We determine the mass ratio of
Nova~Sco~1994 to be 0.419$\pm$0.028 (90 percent confidence), which is 
the most accurate  determination of the binary mass ratio in an X-ray binary. 
This
result combined with the binary mass function and inclination angle gives a
refined black hole mass of 5.99$\pm$0.42$\Msun$ (90 percent confidence).
We also perform simulations which show that, for an F-type secondary star,
the standard rotation profile
with zero and continuum value for the line limb-darkening 
coefficient gives a value for $q$ that brackets the value found using the
full geometrical treatment

\end{abstract}


\section{Introduction}

The determination of binary masses in interacting binaries where only one
component is observable, such as in cataclysmic variables (CVs) or low-mass
X-ray binaries (LMXBs), requires a radial velocity study of the mass
loosing star, a knowledge of the binary inclination angle and mass
ratio. The binary inclination in non-eclipsing binaries is usually
obtained by measuring the ellipsoidal modulation of the late-type star
(e.g. see Wilson \& Devinney 1971 and Shahbaz, Naylor \& Charles 1993) 
and the binary mass ratio is
normally obtained by measuring the rotational velocity of the secondary star 
(Wade \& Horne 1988).  This procedure has been used for many years to determine
binary masses, in particular in the subclass of the LMXBs called the soft
X-ray transients (SXTs), which are characterised by their episodic
outbursts and contain some of the best stellar mass black hole candidates 
(van Paradijs \& McClintock 1995). It is clear that if one wants to
determine accurate binary masses for the SXTs, the uncertainties in
measuring the inclination angle and mass ratio have to be reduced. This
paper concentrates on reducing the inherent uncertainties associated in
determining the binary mass ratio.

Shahbaz (1998) showed that in principle, one could use the shape of the
secondary star's absorption lines to determine the binary system parameters,
such as the inclination angle. However, this requires high resolution echelle
spectroscopy, which for most of the faint SXTs is very difficult, even with
10-m class telescopes.  Due to the faintness of the secondary star in the
SXTs, the secondary star's rotational broadening is usually determined by using
intermediate resolution ($\sim$1\AA) spectroscopy. The secondary stars
typically have $\vsini$ values of 30-100 km/s. With intermediate resolution
spectroscopy the information about the shape of the absorption lines is lost,
the only information that can be extracted is the amount by which the star's
spectrum is broadened.  The procedure commonly used to measure the secondary
star rotational broadening is to compare it with the spectrum of a slowly
rotating template star, observed with the same instrumental configuration which
has been convolved with a limb-darkened standard rotation profile (Gray 1992;
Collins \& Truax 1995  and references within).         
For this one needs to assume a limb-darkening coefficient for
each spectral line. Usually one either adopts 0, i.e. no
limb darkening, or the continuum value (which depends on
the wavelength and the star's effective temperature).  The width of the 
standard
rotation profile is varied until an optimum match is found with the width of
the target spectrum (e.g. see Marsh, Robinson \& Wood 1994).

It should be noted that there are many assumptions inherent in
using this method, primarily due to the use of the standard rotation
profile.  The standard rotation profile assumes that the star is spherical
and that the line profile has the same shape over the entire star. It also
assumes a wavelength dependent limb darkening coefficient and a
homogeneous stellar surface. Firstly, it should be noted that the
secondary star in a CV or LMXB substantially fills its Roche lobe so is
far from being spherical and will thus have distorted line profiles the
extent of which depends on the exact Roche geometry (Shahbaz 1998).
Also, the spectrum of the secondary cannot be described by a single-star 
spectrum, since both temperature and gravity vary over the photosphere due
to the shape of the star's Roche-lobe.  Furthermore, stellar absorption
lines have core limb-darkening coefficients much less than the appropriate
continuum value because the line flux arises from higher regions in the
stellar atmosphere than the continuum flux; the precise value for the line
limb-darkening coefficient depends on many parameters such as the
incidence angle and so requires detailed calculations (Collins \& Truax
1995).  Using zero or the continuum value introduces a systematic
uncertainty of about 14 percent in the determination of $\vsini$ (Welsh
et al., 1995, Shahbaz 1998)  which contributes significantly to the accuracy 
which one can determine the mass ratio and hence the binary masses.

In an attempt to reduce the uncertainties in the binary mass ratio derived
from intermediate spectroscopy using this procedure, we determine the
exact rotationally broadened spectrum from the secondary star in an
interacting binary. Our calculations take into account the shape of the
secondary star's Roche-lobe and the varying temperature and gravity of
different photospheric segments, which are characterised by individual
model atmospheres. By directly comparing the model spectrum with the
observed intermediate resolution spectra of an SXT, we show that we can
accurately determine the binary mass ratio. 
As an example, we determine the binary mass ratio for the well studied
soft X-ray transient, Nova~Sco~1994 (=GRO J1655-40).

%
\begin{table*}
\caption{SXTS with spectroscopic mass ratio measurements}
\label{tble:mratio} 
\begin{center}
\begin{tabular}{lccl}

Object        &  $\vsini$ (km/s)  & $q (=M_2/M_1$) & Reference\\
              &                   &     & \\   
GRO~J0422+32  & 90$^{+22}_{-27}$  & 0.116$^{+0.079}_{-0.071}$  
              & Harlaftis et al., (1999) \\
A0620--00     & 83$\pm$5   & 0.067$\pm$0.01  
              & Marsh, Robinson \& Wood (1994) \\
GS2000+25     & 86$\pm$8   & 0.042$\pm$0.012 
              & Harlaftis, Horne \& Filippenko (1996)\\
Nova~Mus~1994 & 106$\pm$13 & 0.128$^{+0.044}_{0.039}$ 
              & Casares et al., (1997) \\
Cen~X-4       & 43$\pm$6   & 0.17$\pm$0.06   
              & Torres et al., (2002) \\
XTE~J1550-564 & 90$\pm$10  & 0.152$^{+0.048}_{0.042}$ 
              & Orosz et al., 2002 \\
Nova~Sco~1994 & 82.9--94.9 & 0.337--0.436    
              & S99 \\
V4641~Sgr     & 123$\pm$4  & 0.667$\pm$0.356
              & Orosz et al., 2001 \\
V404~Cyg      &	39.1$\pm$1.2 & 0.060$^{+0.004}_{-0.005}$
              & Casares \& Charles (1994) \\
\end{tabular}
\end{center}
\end{table*}
%

\section{Determining the spectrum in interacting binaries}

Although there are many codes that use model atmospheres to determine the
light curves of binary systems (Wilson \& Devinney 1971, Tjemkes, van
Paradijs \& Zuiderwijk 1986, Shahbaz et al., 1993, Orosz \& Hauschildt
2000) there are not many that compute the observed spectrum. Linnell \&
Hubeny (1994, 1996) have a code that computes the synthetic spectra and
light curves for binary stars. However, since they use Kurucz (1979) model
atmospheres to synthesize the spectrum, their code is really only
applicable for systems with relatively hot stellar components i.e.
early-type stars. In LMXBs, the low-mass companions are typically
late-type K stars with effective temperatures less than 5000~K.  Also,
since the temperature varies across the star's surface due to gravity
darkening, there are elements on the star's surface much cooler than the
star's effective temperature, especially near the inner Lagrangian point
($L_1$; Shahbaz 1998). Therefore, in order to compute the observed
spectrum of the late-type secondary star in an LMXB accurately and to make
the code more general for a range of secondary stars, we have written a
code that is similar to the code used in Shahbaz (1998) but with a full
treatment of limb darkening through the addition of synthetic $\sc NextGen$
model atmospheres.

\section{The model}

The equations that determine the basic geometry of an interacting binary
system have been known for a long time and are well understood.  To
determine the light curve or spectrum from the secondary star in a close
binary, one would compute the specific intensity for each element of area
on the star's surface by solving the radiation transfer equation
for a given model photosphere.  By doing this one eliminates the need to use
a limb-darkening law which parameterizes the specific intensity for
different emergent angles, given the monochromatic intensity. However,
since the computation of stellar atmospheres is computationally intensive,
approximations are required to determine the expected light curve or
spectra from an interacting binary system.

The main geometrical part of the model we use is the same as that
described in Shahbaz (1998) and is similar to other standard models for
interacting binaries (Tjemkes et al., 1986, Orosz \& Bailyn 1997).  
Briefly, we model a binary system in which the primary is a compact
invisible object and the secondary a "normal" star. We assume that the
secondary fills its Roche-lobe, is in synchronous rotation and has a
circular orbit. A grid consisting of a series of quadrilaterals of
approximately equal area is set up over the Roche surface. The
grid is defined using a polar coordinate system with $N_\theta$ rings
equally spaced across the star, from the $L_1$ point to
the back of the secondary star (see Orosz \& Bailyn 1997 for details). 
For each ring
of constant angle $\theta$ the number of azimuth points is chosen to try
to keep the area roughly constant. We typically use $N_\theta$=40, which
corresponds to 2048 surface elements.

The parameters that determine the geometry of the system are the binary
inclination $i$, the binary mass ratio $q$ (=$M_2/M_1$, where $M_1$ and
$M_2$ are the masses of the compact object and secondary star
respectively)  and the Roche-lobe filling factor, $f$. Since we assume the
secondary fills its Roche-lobe, a safe assumption in LMXBs, $f$=1.  The
radial velocity semi-amplitude, $K_2$, determines the velocity scale of
the system. The velocity, gravity and temperature for each element varies
across the star due to the shape of the Roche-lobe, the Roche-potential
and gravity darkening respectively. The temperature and gravity for each
element are scaled using the observed effective temperature 
$<T_{\rm eff}>$ and gravity $<\log g>$,
which is determined from the spectral type and class of the secondary
star.

The local gravity is scaled so that the integrated gravity over the
star surface is given by the observed gravity. 
The temperature of each element is calculated using the well-known 
von Zeipel relation (von Zeipel 1924), 
$T \propto g^{\beta}$ where $\beta$ is the gravity darkening 
exponent, and is
scaled so that the integrated luminosity over the stellar surface
matches the observed bolometric luminosity.
The limb angle $\mu=\cos \theta$, is the angle
between the local surface normal and the line of sight to the observer;  
$\mu$=1 at the center of the star and $\mu$=0 at the limb. Each element of
area on the surface of the star is assigned a temperature,
gravity, projected velocity and limb angle $\mu$. For a fixed wavelength,
the specific intensity for a given element of area depends on $\Teff$,
$\log g$ and $\mu$.

\subsection{The synthetic spectra}

To compute the specific intensity at a given wavelength we use the
multi-purpose state-of-the-art stellar atmosphere code $\sc phoenix$,
which produces $\sc NextGen$ model atmospheres, the details of which can
be found in Hauschildt, Baron \& Allard (1997) and Hauschildt et al.,
(1999).  The $\sc NextGen$ models are the most detailed models available
for cool stars and are computed using spherical geometry, rather than the
usual plane-parallel approximation. Since computing $\sc NextGen$ models
are heavily time consuming, we cannot compute spectra for all effective
temperatures, gravities and limb angles.  Therefore,  we precompute spectra 
for set $\Teff$, $\log g$ and $\mu$ combinations.

We compute model spectra with solar abundance in the wavelength  range 6300 to
6800 \AA\ in steps of 0.1 \AA\, $\Teff$ in the range 2400 K to  9800 K in steps
of 200 K and $\log g$ in the range 1.5 to 5.0 in steps of  0.5.  The wavelength
range above was chosen because this region is commonly  used to determine the
secondary star's rotational broadening since it has  many strong Fe and Ca
absorption line features intrinsic to  late-type stars. For a given  $\Teff$
and $\log g$, $\sc phoenix$ computes the specific intensities for 64  different
angles which are chosen by the code based on the structure of the  atmosphere.
For consistency between the models we interpolate the specific  intensities
onto a regular grid with $\mu$ values ranging from 0.05 to 1.0.  For 
accuracy
we use a spline interpolation method (Press et al., 1992) in 
logarithmic  units. Also, we  set the specific intensity to be zero  for 
$\mu <$ 0.05, since at such high limb angles the specific  intensity
is negligible (see Orosz \& Hauschildt 2000 for details).  Finally, to
determine the specific intensity for a given wavelength we perform a 3-D 
logarithmic  interpolation in $\Teff$, $\log g$ and $\mu$ (see section 3.2).

For each element of area visible we compute $\Teff$, $\log g$ and $\mu$.
We then determine the specific intensity at each specified wavelength
and integrate the visible specific intensity values over the Roche surface.
We velocity-correct the model
spectra by first converting the wavelength scale from vacuum to air (Allen
1973) and then bin the spectra onto a uniform wavelength scale. Finally,
given a colour excess we redden the spectrum using the reddening law from
of Seaton (1979). We thus obtain the exact rotationally broadened spectrum
of the secondary star at a given orbital phase. The model parameters that
determine the width and shape of the model absorption lines are $q$, $i$,
$K_2$ and $\beta$; $q$ and $K_2$ determine primarily determine the width
of the lines, whereas $i$ and $\beta$ mainly determine the shape of the
absorption line (also see Shahbaz 1998).

\subsection{The accuracy of the interpolation scheme}

Here we discuss the numerical accuracy of the interpolation method we  employ.
For a fixed wavelength we precompute the specific intensity for set of
($\Teff$, $\log g$, $\mu$) grid points. Given a model combination  of 
($\Teff$, $\log g$, $\mu$)  we perform a 3-D logarithmic interpolation to
determine  the specific intensity.  Orosz \& Hauschildt (2000), showed
that the specific intensity values over a given bandwidth are a relatively 
smooth function
of temperature and gravity. We therefore tried linear and cubic spline
interpolation methods. In contrast to Orosz \& Hauschildt (200), we found 
that linear or cubic spline interpolation was not accurate and robust 
enough, especially towards the  limb of the star. Note that we 
interpolate
absorption lines strengths at a given wavelength, whereas Orosz \& Hauschildt
(2000) interpolate over  a large bandwidth. The effects of absorption lines
that do not vary  smoothly would naturally be smoothed out when integrated over
a large  bandwidth. The interpolation method we use is  based on a modification
of Shepards method for a set of  3D scattered data  points and is described in
Renka (1988).  Basically the method computes a smooth trivariate function using
an least squares fit to the grid points near the interpolant. This function is
then used for the interpolation.

A simple way to test the interpolation scheme is to remove one 
($\Teff$, $\log g$ and $\mu$) grid point from the precomputed intensity grid 
and then calculate its interpolated value. This value can then be compared to 
the original grid point value and gives an indication of the accuracy of the
interpolation method.
We compute a normal spectrum
where we used all the grid points in the intensity table and a  clipped
spectrum where we removed the grid point at $\Teff$=4500~K,  $\log g$=4.5 and
$\mu$=0.10 from the table. We find the interpolation  method to be accurate to
$<$0.5 percent.  To determine to effects of using a relatively coarse grid for
the  Roche-lobe, we computed spectra using two different grids, a standard grid 
with $N_\theta$=40 and a fine grid with  $N_\theta$=60, which corresponds 
to
2048 and 4194 grid points respectively. We found that the two spectra agreed to
within 0.8 percent. Therefore, we believe that our integration and
interpolation schemes  are accurate to better than 1 percent.

%
%
\begin{figure}
\vspace{0.5cm}
\hspace*{-0.5cm}
\psfig{file=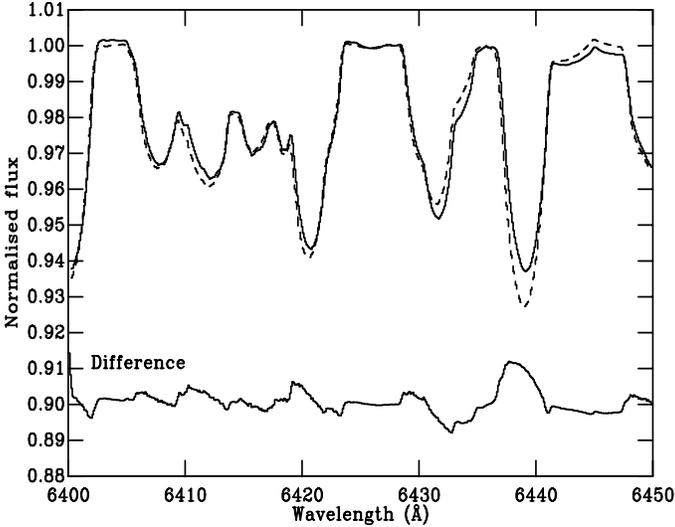,width=10.cm,height=8cm,angle=90} 
\vspace{-1.5cm}
\caption{ 
The effects of using a limb-darkening law to compute  a Roche-lobe model
spectrum. The solid line is the high resolution model spectrum (top) 
where the limb
darkening is  computed properly. The model parameters are
$q$=0.42, $K_2$=215.5\kms, $i$=70$^{\circ}$, $\beta$=0.08, 
$<T_{\rm eff}>$=6400~K and $<\log g>$=4.0).
The dashed line is the high resolution model
spectrum where a linear limb-darkening law has been used with a
limb-darkening coefficient of 0.52.
The bottom spectrum is the residual spectrum, shifted upwards to clarity.
}
\label{fig:limb} 
\end{figure}

\subsection{The effects of limb-darkening}

We compare two high resolution  Roche-lobe model spectra,
one where the limb-darkened spectrum is computed properly and the other
where the spectrum is computed asuming constant limb darkening 
(of 0.52) across the line. 

In both cases, the same Roche-lobe system
parameters are used ($q$=0.42, $K_2$=215.5\kms, $i$=70$^{\circ}$, 
orbital phase=0.35, $\beta$=0.08, $<T_{\rm eff}>$=6400~K and $<\log g>$=4.0).
In Fig.~\ref{fig:limb}, one can see that the spectra are clearly different. 
For the constant limb-darkened spectrum, the limb-darkening is assumed to 
be the same for all the lines. 
Also, a linear or two-parameter law does not 
accurately describe the real limb-darkening behavior, since the intensity 
at the limb of the star is much less (Orosz \& Hauschildt 2000).
Therefore it is not surprising that the shape and strength of the lines 
are not the same, especially for the blended lines.

%
%
\begin{figure*} 
\vspace*{5mm}
\hspace*{2cm}
\psfig{file=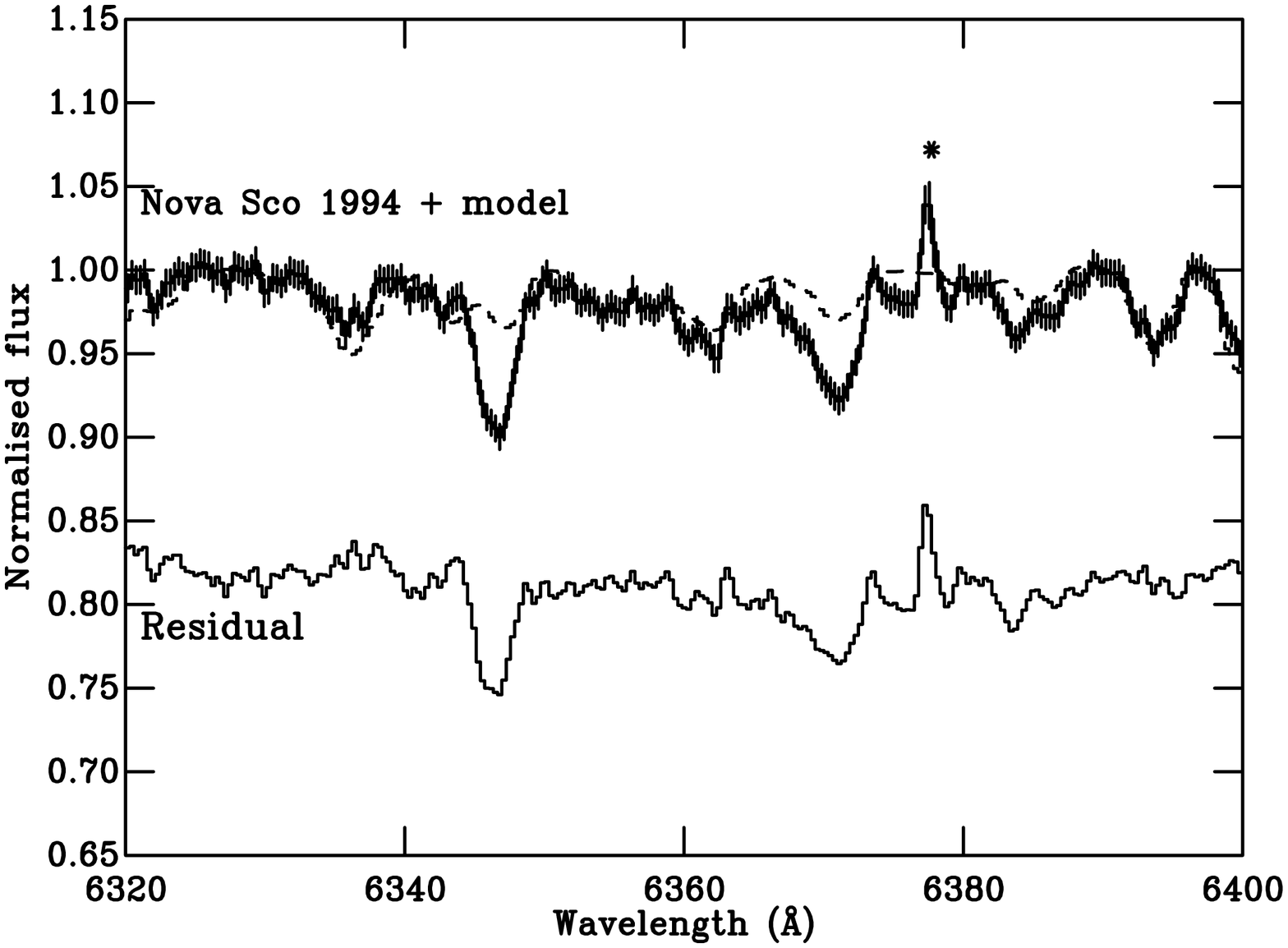,width=17cm,height=9cm,angle=90} 
\hspace*{2cm}
\psfig{file=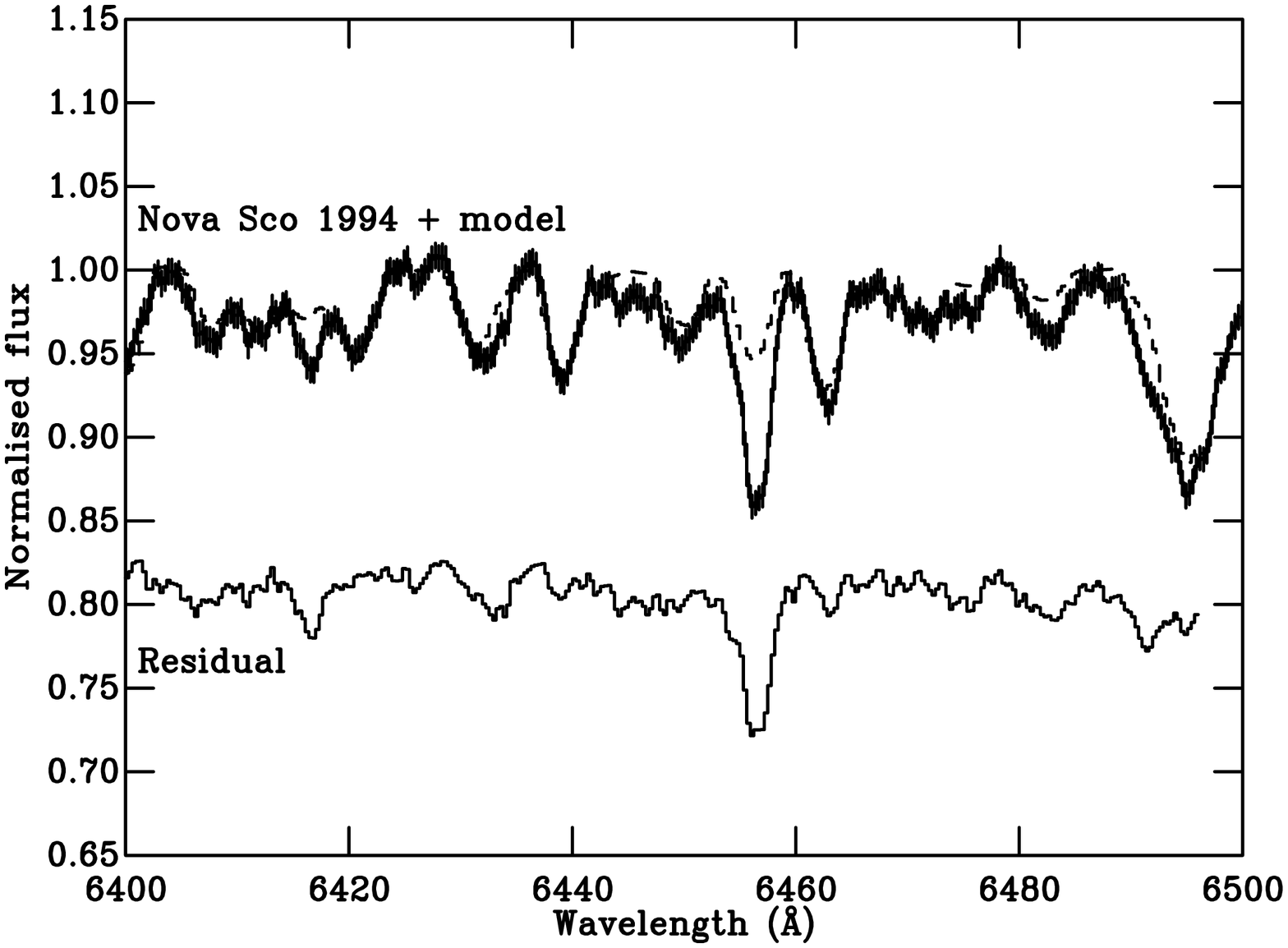,width=17cm,height=9cm,angle=90} 
\vspace*{-10mm}
\caption{ 
The variance-weighted average spectrum of Nova~Sco~1994
(solid line) and the best model spectrum overplotted (dashed line) using 
$q$=0.42, $i$=70$^{\circ}$, 
$\beta$=0.08, $<T_{\rm eff}>$=6336~K  and $<\log g>$=4.0.
For the Nova~Sco~1994 spectrum we also show the error bars on the data.
The residual spectrum of Nova~Sco~1994 after  subtracting
the scaled model spectrum is shown underneath.
The feature marked with an asterisk is residual sky line.
The spectra have been normalised.
All the spectra are in the rest frame of the model.
} 
\label{fig:nsco} 
\end{figure*} 
%

\section{Application to the SXTs: X-Ray Nova Scorpii 1994}
 
Due to the faintness of the secondary star in most of the quiescent SXTs, 
it is usually only possible to obtain intermediate resolution spectroscopy. 
However, this implies  that the shape of the absorption lines cannot be 
used to determine system parameters  (Shahbaz 1998). The only information
which we can use is the  width of the absorption lines, which is related to
$q$.  To date there are 9 measurements of  the binary mass ratio for SXTs
(see Table 1), which were all obtained by measuring the  secondary star's
rotational broadening  and using the standard limb-darkened  rotation profile. 
If we are to determine accurate masses for the binary components in the SXTs,
we have to try and reduce all known sources of uncertainties.  It is  possible
to eliminate the uncertainties in the binary mass ratio obtained by  assuming
the standard rotation profile, by determining  the true limb-darkened 
rotationally broadened spectrum of the secondary star. In this section, we 
outline  this method, using the SXT Nova~Sco~1994 as an example.

\subsection{Previous work}

The SXT Nova~Sco~1994 has been extensively  studied over the past years at
both  X-rays and optical wavelengths [see  Shahbaz   et al., 1999 (hereafter
S99),  Orosz \& Bailyn 1997 and references within].  Optical photometry  taken
during X-ray quiescence  ($L_{\rm X}<10^{-3}L_{\rm opt}$) revealed  the orbital
period $P_{\rm orb}$=2.62168 days (Orosz \& Bailyn 1997).   Through modeling
the  quiescent optical light curves, the binary inclination  was measured to be
70$^{\circ}$ (Orosz \& Bailyn 1997). 
Greene, Bailyn \& Orosz (2001) obtained the same result by modeling
the optical and
infrared  quiescent light curves simultaneously with an upgraded  code. 
Spectroscopic observations allowed
the radial velocity semi-amplitude to be measured $K_2$=215.5$\pm$2.45 km/s,
which implies a binary mass function of 2.73$\pm$0.09$\Msun$ (S99).  The
spectral type was determined to be F3--F6 (Orosz \& Bailyn 1997), and  later
refined to be F6 (S99).

Since the companion star fills its Roche-lobe and is in  synchronous rotation
with the binary motion, $\vsini$ combined with $K_2$  gives a direct
measurement of $q$  (Horne, Wade \& Szkody 1986)

\begin{equation}
v\sin~i = K_{2} (1 + q ) \frac{ 0.49q^{2/3} }{ 0.6q^{2/3} + \ln(1+q)^{1/3} }
~~~~~~~{\rm km~s^{-1}}
\end{equation}

\noindent
S99 used the standard
technique to estimate the rotational broadening of the  secondary star. Using
optical spectra with a velocity resolution of 38.6 \kms, they compared the
average observed spectrum of  Nova~Sco~1994 with field template stars
broadened using the standard rotation profile.  
For limb-darkening 
coefficients of 0.0 and 0.52, they obtained $q$=0.36 and $q$=0.40
respectively. However, it should be noted that the template star they used has
an intrinsic rotational broadening of $\sim$30 \kms (see SIMBAD database),
which implies that the $\vsini$ values were underestimated. In
order to determine $\vsini$ for Nova~Sco~1994,   we repeated the analysis
described in S99, but using the F6$\sc iii $ template star HR5769 
($\vsini$$<$5 \kms; SIMBAD database) and the spectral region 
6380\AA--6500\AA.
For limb-darkening coefficients of 0.0 and 0.52, we
obtain $\vsini$ values  of  89.6$\pm$2.8\kms 94.8$\pm$3.0\kms respectively,
which correspond to a mass  ratio of 0.385$\pm$0.023 and 0.427$\pm$0.024
respectively (90 percent uncertainty). We find that the spectrum of the
secondary star is not veiled ($f_v$=1.00$\pm$0.05). These results are
consistent with the  $\vsini$ and veiling values found by  Israelian et al.,
(1999) using high resolution spectra. Note that the difference between the
$\vsini$ values for Nova~Sco~1994  obtained using HR2927 (S99) and HR5769 is
$\sim$20\kms. This difference is the rotational velocity of HR2927 and agrees
well with the value in the  SIMBAD database.

In order to obtain a more accurate determination of $q$, which does not depend 
on assumption about the rotation profile and limb-darkening coefficients,  we
use the model described in section 3 to compute the exact rotationally 
broadened model spectrum, which can then be directly compared with the  
average observed spectrum of Nova~Sco~1994 published in S99. Since the Nova~Sco~1994 
spectra have a resolution of 38.6 \kms, we cannot use the shape of the
absorption  lines to determine system  parameters such as $i$ or $\beta$
(Shahbaz 1998).  The only information which we can use is the width of the
absorption lines,  which is related to $q$.

\subsection{Model parameters}

In our model, the temperature distribution across the secondary star is 
described by a gravity darkening law. For stars with radiative atmospheres the
exponent ($\beta$) for the gravity darkening law is 0.25 (von Zeipel 1924), 
whereas if the atmosphere of  the star is convective, the  exponent is 0.08
(Lucy 1967). The spectral type of the secondary star in Nova~Sco~1994
(F6) is on the  boundary between hot stars with radiative envelopes and cool
stars with  convective envelopes. Claret (2000) has shown that the value  for
$\beta$ is a function of the star's effective temperature.
For Nova~Sco~1994 with $<T_{\rm eff}>$=6336~K, $\beta$=0.08.
However, to be conservative in our model computations, we use $\beta$=0.08
and 0.25.

We fix $i$ to be  70$^{\circ}$ (Greene et al., 2001) and
assume the secondary to be a F6$\sc iv$ (S99) with an 
$<T_{\rm eff}>$=6336~K and $<\log g$=4.0 (Gray 1992). 
For $\beta$=0.08 the temperature and gravity across the secondary star 
ranges from  5120~K to 6490~K and $\log g$=2.8 to 4.1 respectively.
The temperature distribution for
$\beta$=0.25 is much wider  and flatter (3230~K to 6780~K) 
and there is no difference in the range of $\log g$.
	
For a compromise between speed and resolution, we choose the number of grid 
points to be such that the  maximum radial velocity difference between any two
adjacent elements is less than the instrumental resolution. We choose the 
total number of surface elements to be 2048  ($N_\theta$=40; see section 3), 
which with the binary parameters of Nova~Sco~1994, gives a maximum
radial  velocity difference of less than 10\kms. 

%
%
\begin{figure}
\vspace*{0.5cm}
\hspace*{-0.5cm}
\psfig{file=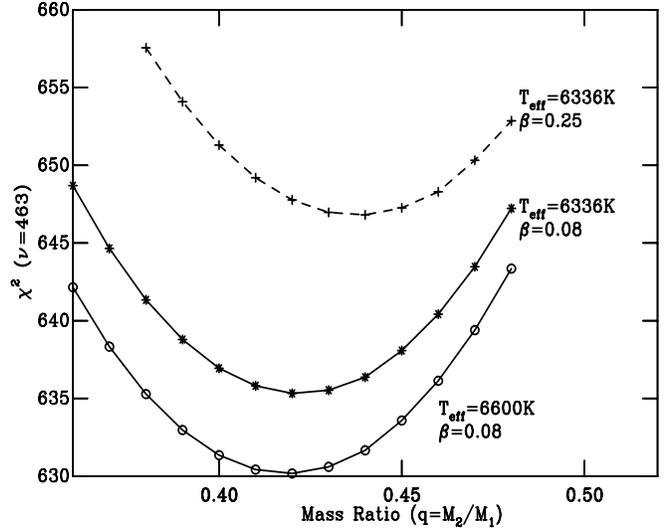,width=10cm,height=8cm,angle=90} 
\vspace{-15mm}
\caption{ The
$\chi^2$ values for the optimal subtraction as a function $q$. 
The stars and crosses show the $\chi^2$ values for fits using 
$\beta$=0.08 and 0.25 respectively with $<T_{\rm eff}>$=6336~K.
The circles show the $\chi^2$ values for fits using 
$<T_{\rm eff}>$=6600~K and $\beta$=0.08. The minimum  $\chi^2$ is 
618.6 and the curve has been shifted upwards.
}
\label{fig:chi2} 
\end{figure}
%

%
%
\begin{figure*}
\vspace*{5mm}
\hspace*{2cm}
\psfig{file=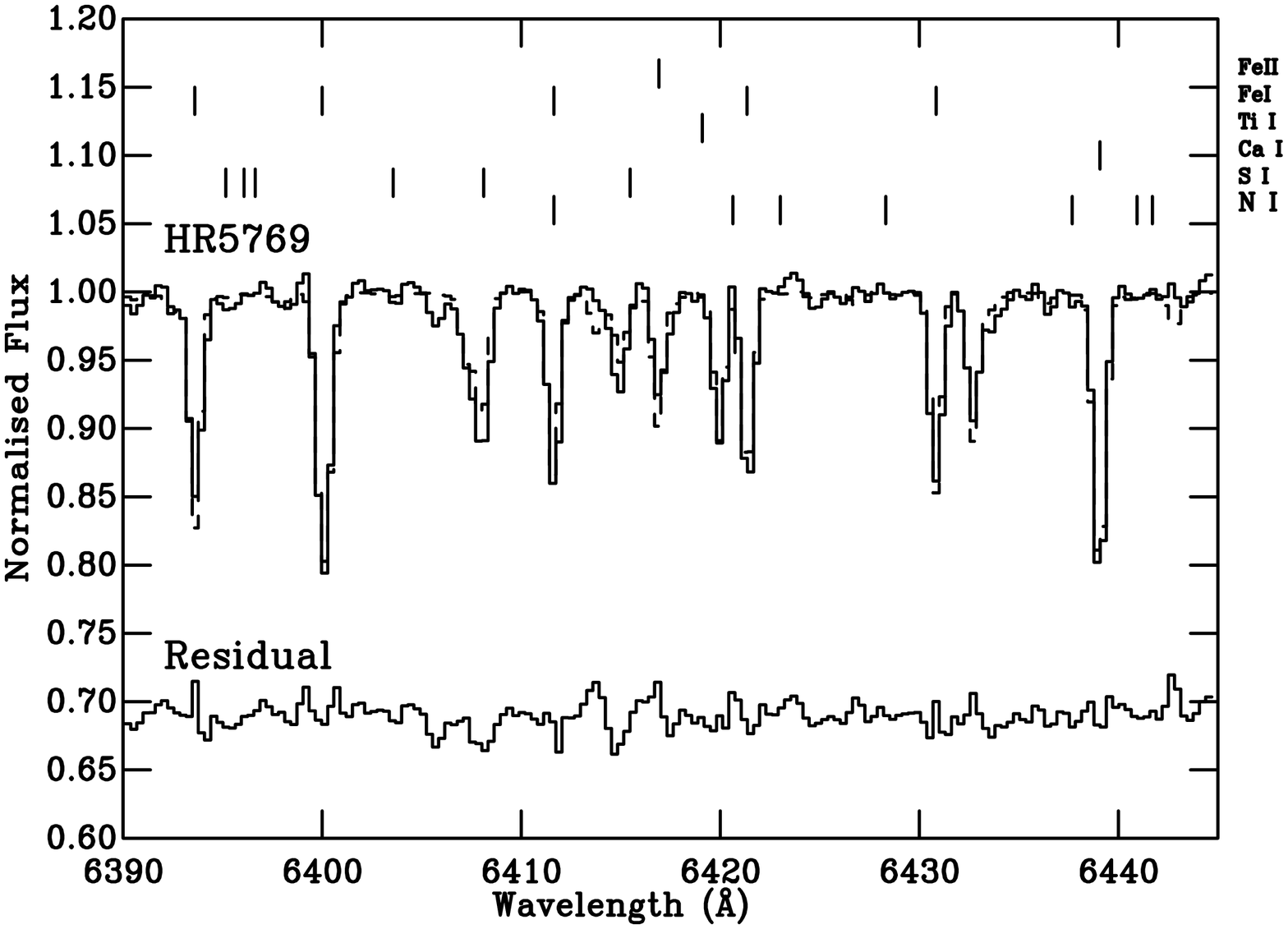,width=17cm,height=9cm,angle=90}
\hspace*{2cm}
\psfig{file=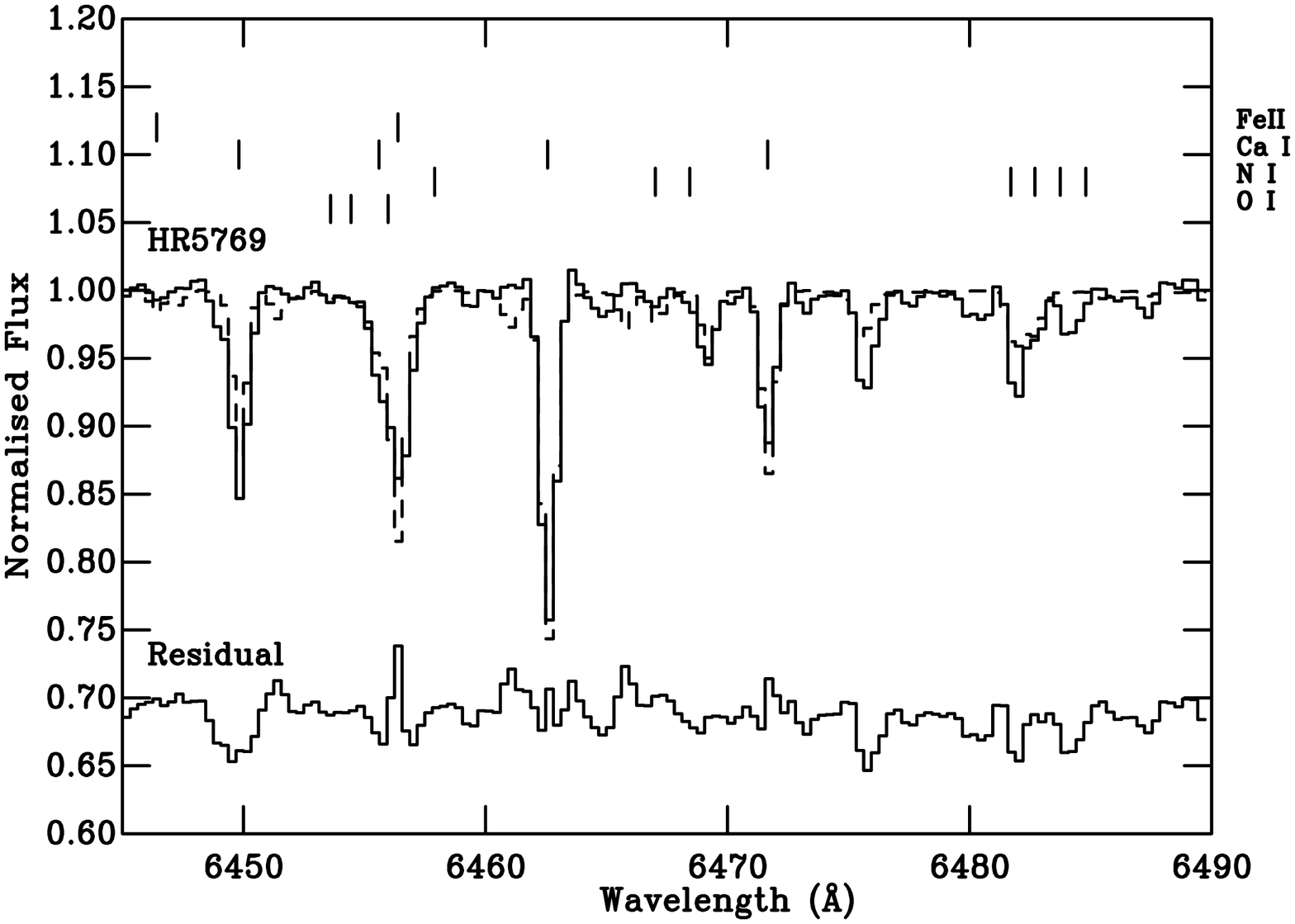,width=17cm,height=9cm,angle=90}
\vspace*{-10mm}
\caption{ 
The template F6$\sc iv$ star HR5769 (solid line) and the   $\sc NextGen$
model spectrum (dashed line) computed using $\Teff$=6400~K and $\log g$=3.7 
and degraded to match the resolution of the template star spectrum. 
The residual   spectrum is also shown at the bottom.
The spectra have been normalised.
The wavelength position of the atomic Fe lines and the 
$\alpha$-elements are also shown.
}
\label{fig:hr5769}  
\end{figure*} 
%

\subsection{The binary mass ratio }

Our analysis was performed in the same way as by S99, the only difference
being  that we compute broadened spectra using our model which depends on $q$, 
instead of using broadened template stars. For a given value for $q$ we  first
computed the model spectrum in the wavelength range 6300~\AA\ to 6500~\AA.  
This region was chosen since it is similar to the region used in the  analysis
of S99 and has many relatively isolated absorption line features. We then 
broadened the model spectrum using a Gaussian  function with a width 38.6 km/s
to match the instrumental resolution of the  Nova~Sco~1994 data. We determined
the velocity shift of the average  observed spectrum of Nova~Sco~1994 with
respect to the model spectrum using the method of cross-correlation (Tonry \&
Davis 1979). Prior to the  cross-correlation all the spectra were rebinned onto
a logarithmic  wavelength scale with a pixel size of 14.5 \kms. 

In order to compare our  model spectrum with the   average observed
Nova~Sco~1994 spectrum, one should compute the  model spectra in the same way
as the Nova~Sco~1994 spectra were obtained and averaged.    The Nova~Sco~1994 
average spectrum consisted of 25 velocity-corrected spectra taken  during 2
nights covering orbital phases  near  0.35 and 0.85.   Ideally one should
compute model spectra at the same orbital  phases and allow for the orbital
smearing due to the exposure time. However,  the resolution of the data is such
that any information about the shape of the absorption lines is  smeared out 
(Shahbaz 1998). Therefore, we compute a single model spectrum at phase 0.35 
and assume it is the same for all orbital phases.  

To allow for the smearing of the absorption lines due to the  motion of the
secondary star during the length of each exposure at each  orbital phase, we
convolved the model spectrum with a rectangular function  with the appropriate
velocity width which is given by 
$2 \pi K_2 \cos (2 \pi \phi) / P_{\bf orb}$ \kms.  
We first normalise using a  continuum spline fit and then average the
25 model spectra. Finally, to compare the average model spectrum with
the  average observed spectrum of Nova~Sco~1994 we used the method 
described in Marsh et al., (1994). We subtract a constant, $f_v$, multiplied 
by the model spectrum from the average observed Nova~Sco~1994 spectrum. 
The constant represents the fractional contribution of the template star
spectrum to the total flux density. Therefore 1-$f_v$ is the 
veiling factor and for no veiling $f_v$=1.0.
The optimum value for $f_v$ was obtained by minimising the $\chi^2$ statistic.
We performed the above analysis for $q$ in the range 0.32 to 0.4`52 in steps  of
0.01 and in the wavelength range 6310--6490\AA. 
The residual sky feature at 6376\AA\ was masked.
We obtain a minimum  $\chi^2$ of 1729.1 ($\nu$=530) at $q$=0.42 
In Fig.~\ref{fig:nsco} we show the best model spectrum with $q$=0.42,  the
average observed spectrum of Nova~Sco~1994 and the residual spectrum. 
The poor fit is primarily due to the featuress at 6346\AA\ and 6456\AA\ 
contributing significantly to the poor $\chi^2$.
Masking out these features we obtain a minimum $\chi^2$ of 635.3 
($\nu$=463, $\beta$=0.08).  
Fig.~\ref{fig:chi2} shows the $\chi^2$ plot for different values of $q$. 
For the models with $\beta$=0.08 (circles) the  minimum $\chi^2$ is
obtained at $q$=0.422$\pm$0.028 with $f_v$=0.88$\pm$0.04 
(90 percent uncertainty).
Using a gravity darkening exponent of $\beta$=0.25 (crosses), we obtain a
significantly higher $\chi^2$; the minimum  $\chi^2$ is 646.6 at
$q$=0.439$\pm$0.028 with $f_v$=0.89$\pm$0.04 (90 percent uncertainty).

From optical spectroscopy, S99 and Israelian et al., (1999) showed that  there
is no evidence that the light from the secondary star is veiled (see also
section 4.1). From our fits using  $<T_{\rm eff}>$=6336~K and $\beta$=0.08 
we obtain $f_v$=0.88, i.e. the model predicts deeper absorption lines 
compared to the data.  Since our analysis mainly uses Fe absorption lines, 
which are stronger in  cooler stars, we expect by using a hotter secondary 
star in the model we will obtain a higher value for $f_v$, 
Using $<T_{\rm eff}>$=6600~K and $\beta$=0.08 we obtain a minimum  $\chi^2$
of 618.6 with $f_v$=1.00, i.e. the  secondary star contributes all the flux.
The minimum $\chi^2$ is at $q$=0.419$\pm$0.028 (90 percent uncertainty). 
A star with $<T_{\rm eff}>$=6600~K would have an F3 spectral type
(Gray 1992),  which  is consistent with the observed spectral type (F3--F6;
Orosz \& Bailyn 1997  and S99).

It should be noted that, although the minimum $\chi^2$  obtained using  
$<T_{\rm eff}>$=6600~K is significantly lower than the minimum $\chi^2$ 
obtained using  $<T_{\rm eff}>$=6336~K, it should be noted that the  shape and
position of the minimum $\chi^2$ does not change; the $\chi^2$  values are only
offset by a constant amount. 
Also, for a fixed  $<T_{\rm eff}>$ the fits  using $\beta$=0.25 give higher
values for $f_v$ compared to the fits using $\beta$=0.08 (see above). Therefore the
determination of $q$ and its uncertainty does not depend  on the choice of
$<T_{\rm eff}>$.
If we use an inclination of 68$^{\circ}$ or 72$^{\circ}$, or
if we change $K_2$ by 4 \kms (the 90
percent limits on $K_2$; see S99) the  derived value for $q$ does not change;
the  minimum $\chi^2$ increases by less than 1.
Note that the uncertainties in the temperature/spectral type and
in $\beta$ are  of the same order as that introduced by using the standard
rotational profile.

%
\begin{table}
\caption{Summary of model fits. 90 percent uncertainties are given}
\label{tble:fits} 
\begin{center}
\begin{tabular}{lll}

Nova~Sco~1994 & Standard profile &                        \\
&          u=0.00  &  $q$=0.385$\pm$0.023  \\
&          u=0.52  &  $q$=0.427$\pm$0.024  \\
	 & 	   			  \\
Nova~Sco~1994 & Roche profile	 & 	   	          \\
&  $<T_{\rm eff}>$=6336~K, $\beta$=0.08  & $q$=0.422$\pm$0.028 \\
&  $<T_{\rm eff}>$=6600~K, $\beta$=0.08  & $q$=0.419$\pm$0.028 \\
	 & 	   			  \\
Simulation  &     &   \\
($q$=0.42) & Standard profile &             \\
& u=0.00   & $q$=0.392$\pm$0.006 \\
& u=0.52   & $q$=0.434$\pm$0.006 \\
\end{tabular}
\end{center}
\end{table}
%

\subsubsection{Overabundant $\alpha$-elements}

The poor minimum $\chi^2$ we originally obtained  before masking  can be
explained either in  terms of the  $\sc NextGen$ models failing to reproduce
the strength of some of the absorption line features used in the optimal
subtraction or that the Nova~Sco~1994 spectrum contains  some overabundant
elements. 
To see if this mismatch is due to the former, we compare a $\sc NextGen$ model
spectrum with a F6$\sc iv$ template  star spectrum. The $\sc NextGen$ model
spectrum was computed  using $\Teff$=6400~K  and $\log g$=4.0, values
appropriate for the  template  star HR5769.
The model parameters were $f$=0.1 (i.e. for a spherical star), 
$K_2$=0.0\kms and $\beta$=0. From 
Fig.~\ref{fig:hr5769}  it can be seen that we have confidence that our code
matches the absorption line features: the scatter in the residual of the lines 
is less than 2 percent.

\begin{figure*}
\vspace*{5mm}
\hspace*{2cm}
\psfig{file=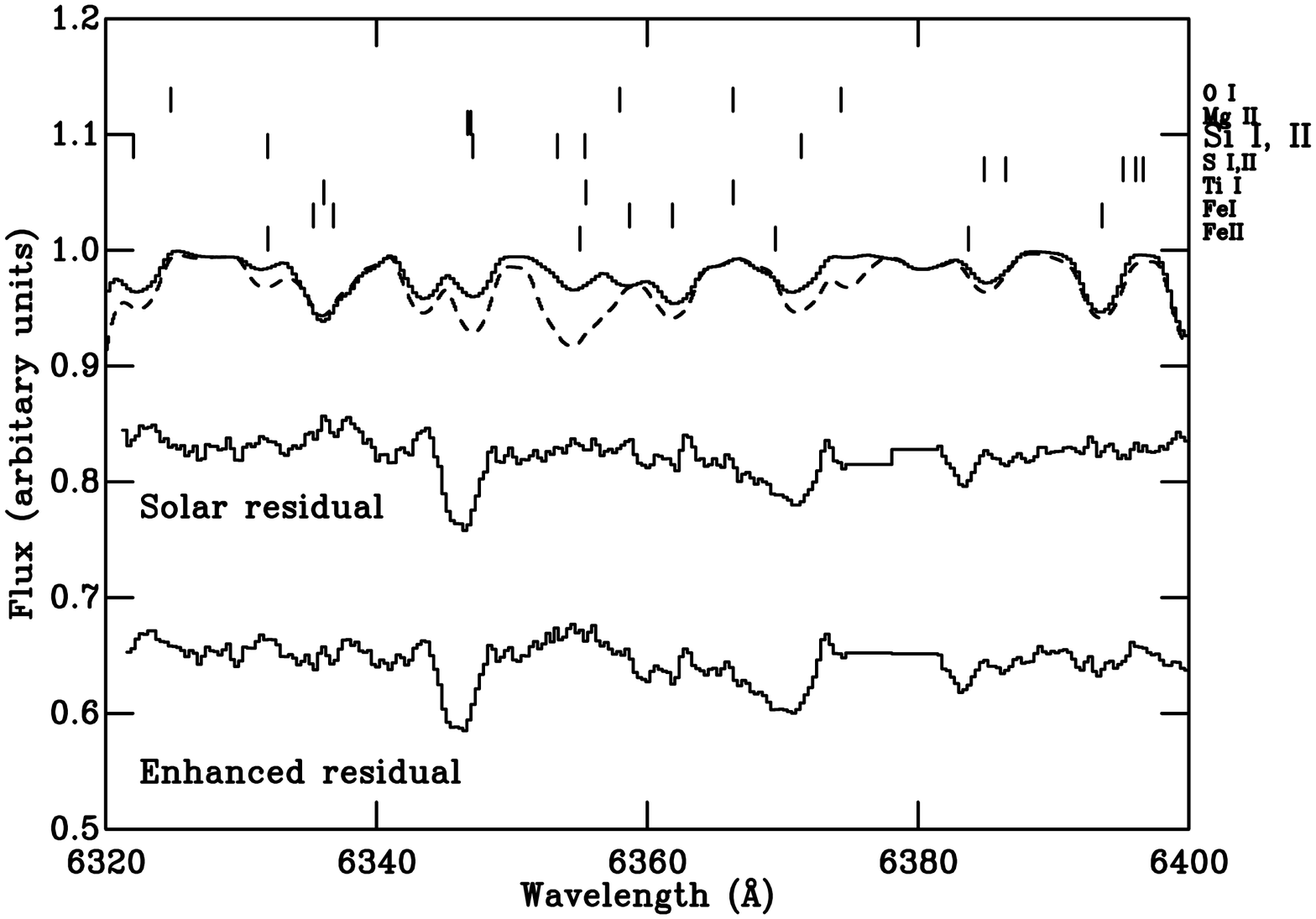,width=16cm,height=9cm,angle=90}
\hspace*{2cm}
\psfig{file=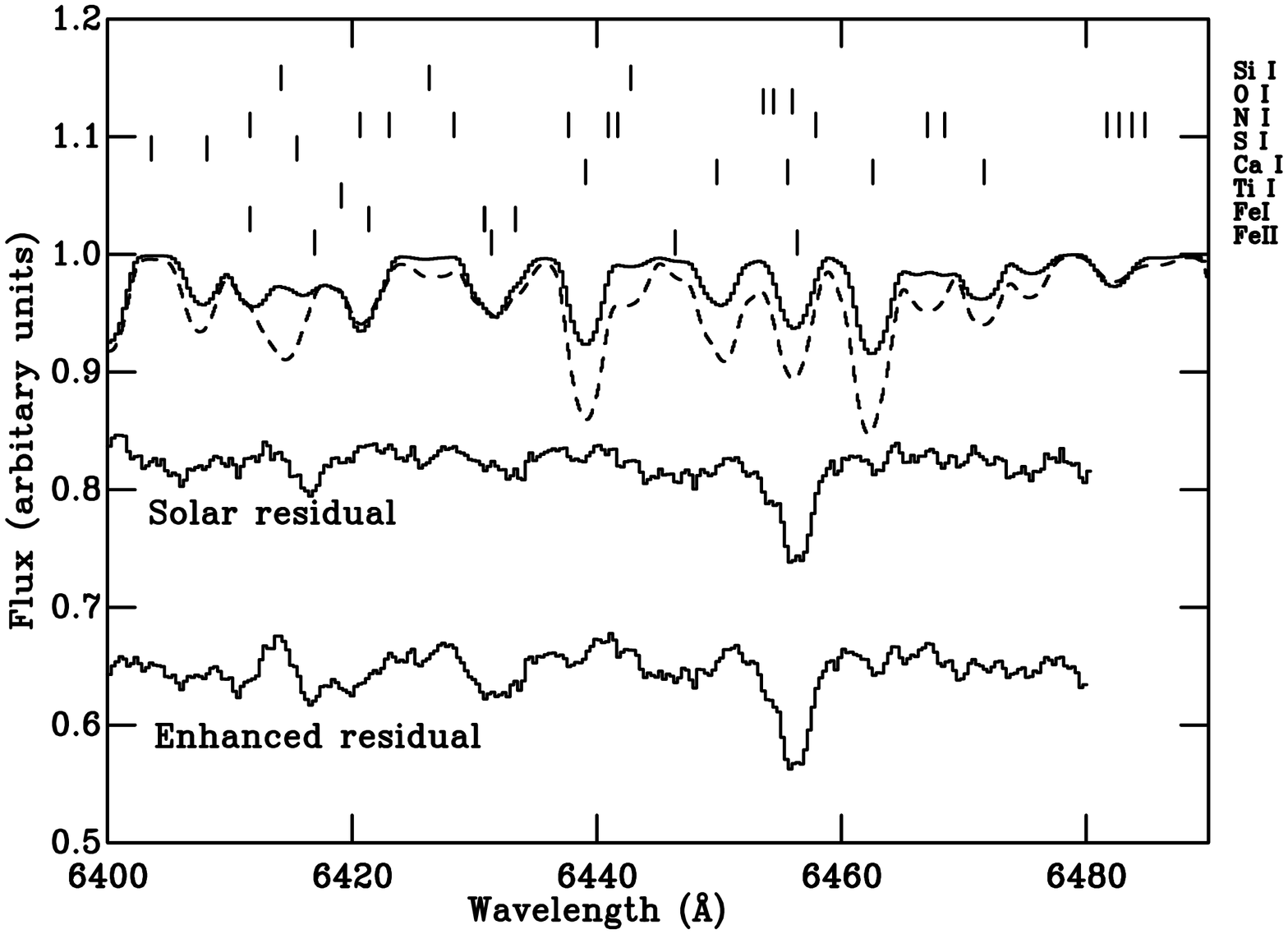,width=16cm,height=9cm,angle=90}
\vspace*{-10mm}
\caption{ 
A comparison between a solar and non-solar LTE spectrum. The LTE synthetic
spectrum were computed  with  $<T_{\rm eff}>$=6400~K and $\log g$=4.0.  The
solid line shows the LTE spectrum with solar abundance.  The dashed lines shows
the LTE spectrum where the  $\alpha$-elements have abundances of  [O/H]=1.0,
[S/H]=0.75, [Mg/H]=0.9, [Si/H]=0.9, [Ti/H]=0.9, [N/H]=0.45 (Israelian et al.,
1999) and [Ca/H]=1.0. 
The LTE model spectra have been  broadened by 95\kms.
The residual spectrum after subtracting the solar (middle) and 
enhanced-abundance (bottom) model spectrum from the
observed Nova~Sco~1994 spectrum are also shown.
}
\label{fig:alpha}  
\end{figure*}

Israelian et al., (1999) have measured the metal abundances of the secondary 
star in Nova~Sco~1994 and find [Fe/H]=0.1$\pm$0.2 which is consistent with 
being solar. They also found an overabundance in the $\alpha$-elements (O, Ca,
S, Mg, Si, Ti and N) compared to solar values. Since the spectral region we 
use in the optimal subtraction is dominated by Fe lines the use of synthetic 
models with solar abundances seems correct. We would then naturally expect  the
few $\alpha$-element absorption line features present in the spectral  region
used for our analysis to be overabundant and hence contributes  significantly
to the poor $\chi^2$.
To determine the qualitative effects of the $\alpha$-elements abundances,  
we compare a solar abundance spectrum and a spectrum
where the $\alpha$-element abundances have been increased, with the observed 
Nova~Sco~1994. (Fig.~\ref{fig:alpha}). The abundances of the latter spectrum were 
increased by [O/H]=1.0, [S/H]=0.75, [Mg/H]=0.9, [Si/H]=0.9, [Ti/H]=0.9 
and [N/H]=0.45 (Israelian et al., 1999). As Israelian et al., (1999)
did not determine the Ca abundance, we also increased the Ca abundance by
[Ca/H]=1.0 to see if the mismatched feature at 6456\AA\ is due to Ca.
Both spectra were computed using $\Teff$=6400~K  and $\log g$=4.0 and were
broadened  by $\vsini$=95~\kms using the standard rotation profile (u=0.52). 
Subtracting solar- and enhanced-abundance spectra from the observed  
Nova~Sco~1994
spectrum, we obtain $\chi^{2}_{\nu}$ values of 2.95 and 4.15 respectively.
The residual solar- and enhanced-abundance spectra are shown in 
Fig.~\ref{fig:alpha}.
The mismatched feature at 6436\AA\ contain the  $\alpha$-elements
Mg$\sc ii$~6346.742~\AA\ and S$\sc i$~6415.48~\AA, whereas the 
feature at 6456\AA\ contains Ca$\sc i$~6455.598~\AA\ and 
O$\sc i$~6455.977~\AA\, but note that  Fe$\sc ii$~6456.38~\AA\ is also present.

While our spectra seem to show that some of the  mismatched features are due to
the $\alpha$-elements being being overabundant, there are also a few features
that  the non-solar spectrum predicts to be overenhanced.   To some extent this
explains why the $\chi^{2}_{\nu}$ for the enhanced spectrum is higher  than the
$\chi^{2}_{\nu}$ for the solar spectrum. Given the poor resolution of the
spectra, it is difficult to determine  with certainty if the mismatched 
features are entirely due to the $\alpha$-elements being overabundant.  Only a
more detailed abundance analysis using high resolution spectra can prove this
convincingly.

%
%
%

\section{Discussion}

\subsection{The mass ratio and black hole mass }

Beer \& Podsiadlowski (2002) determined $q$  by fitting the optical  lights
curves of Nova~Sco~1994 in a self consistent manner,  without any assumptions
about the distance or interstellar reddening. They obtained
$q$=0.256$\pm$0.039  which is significantly lower than the values obtained by
S99 (see also section 4.1) and Greene et al., (2001). They argue that the high
mass ratio obtained  by Greene et al., (2001) is a consequence of neglecting
the accretion disc   when modeling the  optical and infrared light curve. 
Furthermore, in order to explain the disagreement in $q$, they suggested that
the  secondary star's rotational broadening  measured by S99 was overestimated.
[Note that  the revised determination of $q$ (see section 4.1) does not affect 
the argument presented by Beer \& Podsiadlowski (2002)]. They comment that
since the spectra in S99 were taken at quadratures near orbital phases 0.25 and
0.75 (orbital phase 0.0 is defined as superior conjunction of  the secondary 
star) the effects of orbital smearing would be the greatest.  The orbital
smearing would add a linear amount of  3.5 \kms  to the rotational broadening.
As S99 did not take into account the effects of orbital smearing, the
rotational broadening value obtained by S99 would be overestimated.   

Firstly, we would like to point out that the effects of orbital smearing would
be  the least because the velocity of the secondary star is  least  at the
quadratures. Secondly,  because the velocity width of the rectangular smearing
profile is much smaller than the  rotational broadening, the effects of orbital
smearing is very small. To examine the effects of orbital smearing on the
rotational broadening we performed a simple simulation where  we computed a
Gaussian function with  FWHM=100\kms, appropriate for the observed $\vsini$ 
(see Royer et al., 2002 for a rule of thumb calibration) and then smeared it
with a rectangular profile with a width  of 7\kms. We find that the width of
the Gaussian function before and  after the smearing is the same to within 0.2
percent, therefore, the effects of orbital smearing are negligible.
Also, it should be noted that Beer \& Podsiadlowski (2002) obtained 
$\Teff$=6150~K which is significantly different to what we obtain (see
section 4.3). 
Given the systematic uncertainties in modeling the light curves 
(e.g. the binary inclination is correlated with $\Teff$) 
it is difficult to explain the difference in $q$ obtained by 
Beer \& Podsiadlowski (2002).

Using our value for the binary mass ratio, we can determine the mass of the 
binary components from the mass function equation

\begin{equation}
f(M)=\frac{ P K_2^3 }{ 2\pi G } = \frac{ M_1 \sin^3 i }{(1+q)^2}
~~~~~{\rm M_{\odot}}
\end{equation}

\noindent
With f(M)=2.73$\pm$0.15~$\Msun$ (S99), $q$=0.419$\pm$0.028 (see section 4.3) 
and $i$=70.2$\pm$1.6$^{\circ}$ (Greene et al., 2001), we obtain
$M_1$= 6.59$\pm$0.45~$\Msun$ and $M_2$=2.76$\pm$0.33~$\Msun$. 
The 90 percent uncertainites were obtained using a Monte Carlo procedure.
Although the the absolute value for the binary mass ratio obtained 
by Greene et al., (2001) and in this paper are different,
the uncertainties in the mass ratio are such that the black hole mass we 
obtain is consistent with that obtained by Greene et al., (2001).

\subsection{Roche-lobe versus standard rotation profile}

In S99 we determined $q$ for Nova~Sco~1994  by comparing the
average observed spectrum of Nova~Sco~1994  with the  spectrum of a template
star which had been convolved with the standard rotation profile. 
The standard  rotation profile
requires a limb-darkening coefficient which we assume to be the  value
appropriate for the continuum i.e. we assume that the radiation in the
lines is the same as
for the continuum.  Since the absorption lines in early type stars will have
core limb-darkening coefficients much less than the values appropriate for the
continuum (Collins \& Truax 1995), in order to be conservative  we also used a
limb-darkening coefficient of zero.  Using a non-rotating template star with
line limb-darkening coefficients of zero and 0.52 (note that 0.52 is 
actually the continuum value), we obtained mass ratio's of
$q$=0.385 and 0.427 respectively (see section 4.1).
Using the Roche-lobe model we obtained $q$=0.419 (see section 4.3).
As one can see, the Roche-model gives a mass ratio that lies in between
the values obtained using limb-darkening coefficients of 
zero and 0.52 (continuum) and the standard rotation profile.

To prove this, we simulate  a spectrum and then try to
determine $\vsini$ and hence $q$ in the same way as we did in S99 (see section
4.3) and using the X-ray binary model (section 4.1). We first perform the same
analysis as in S99 to determine $\vsini$, i.e. we compare broadened versions of
the non-rotating  template star HR5769 with the simulated  average spectrum, 
using the  standard  rotation profile with a continuum  limb-darkening
coefficients of 0.0 and 0.52. The  high quality (S/N=500) model spectrum was
computed using  $q$=0.42, $K_2$=215.5\kms,  $i$=70$^{\circ}$, $\beta$=0.08,
$<T_{\rm eff}>$=6400~K and $<\log g>$=4.0.  For limb-darkening coefficients of
0.0 and 0.52 we obtain  mass ratio's of $q$=0.392 and 0.434
respectively  (90 percent uncertainty of 0.006). Using the binary model, we find that we
recover the exact model $q$ value. These simulations
confirm that, for the case of an F-type secondary star,  the standard rotation
profile with zero and continuum value for the line limb-darkening  coefficient
gives a value for $q$ that brackets the value found using the X-ray binary
model.

The limb-darkening coefficient varies across the absorption having the
continuum value in the wings to a reduced value in the line core,
where  the exact value for the line limb-darkening coefficient depends
on the atomic species of the line. Since most of the absorption in the
rotationally broadened line comes from the line core, the value for
the limb-darkening in the line core will dominate the shape of the
line (Collins \& Truax 1995).  
The fact that we obtain a better estimate for $q$ using a limb-darkening 
coefficient slightly less than the continuum value 
suggests that the average line limb-darkening coefficient for 
the strongest absorption lines used in the  analysis (i.e. Fe) is close
to the continuum value.  
It is clear that only detailed line profile calculations on
individual lines can determine the true  absorption line 
limb-darkening coefficient.

\section{Conclusions}

We present a model that predicts the  spectrum
of the secondary in an interacting binary. The model uses synthetic $\sc
NextGen$ model spectra which are incorporated into the secondary star's
Roche geometry. As a result, we determine the exact rotationally broadened
spectrum of the secondary which does not depend on assumptions about the
rotation profile and limb-darkening coefficients.

As an example, we determine the mass ratio for the SXT Nova~Sco~1994. We
use our model to compute the  rotationally broadened model spectrum,
which we compare directly with the observed intermediate resolution
spectrum of Nova~Sco~1994.  In order to match the width of the absorption
lines and the lack of veiling in the quiescent optical spectrum, our model
requires $q$=0.419$\pm$0.028 (90 percent uncertainty), and a secondary star
with $<T_{\rm eff}>$=6600~K (an F3 star) which is consistent with the
observed spectral type. For the case of an F-type secondary star, 
we find that the 
standard rotation profile with zero and continuum value for the line
limb-darkening  coefficient gives a value for $q$ that brackets the value found
using the full geometrical treatment.

\section*{Acknowledgments}

I am deeply grateful to Peter Hauschildt for computing the $\sc NextGen$ 
synthetic spectra. 
I would also like to thank the referee for the careful reading of the 
manuscript and for the extremely useful comments that improved the paper.
TS was supported by an EC Marie Curie Fellowship HP-MC-CT-199900297.
 
{}


\begin{thebibliography}{}

\bibitem{} 
Allen C.W., 1973, Astrophysical Quantities, Athalone Press, London.

\bibitem{} 
Beer M., Podsiadlowski Ph., 2002, MNRAS, 331, 351

\bibitem{} 
Casares J, Charles P.A., 1994, MNRAS, 271, L5

\bibitem{} 
Casares J., Martin E.L., Charles P.A., Molaro P., Rebolo R., 
1997, NewA, 1, 299 

\bibitem{} 
Claret A., 2000, A\&A, 359, 289

\bibitem{} 
Collins~$\sc ii$ G.W., Truax R.J., 1995, ApJ, 439, 860

\bibitem{} 
Gray D.F., 1992, in The observation and Analysis of Stellar
Photospheres,  Cambridge Astrophys. Ser. 20, Cambridge University Press,
Cambridge

\bibitem{} 
Harlaftis E., Horne K., Filippenko A.V., 1996, PASP, 108, 762 

\bibitem{} 
Harlaftis E., Collier S., Horne K., Filippenko A.V., 1999, A\&A, 341, 491 

\bibitem{} 
Hauschildt P.H., Allard F., Baron E., 1997, ApJ, 512, 377

\bibitem{} 
Hauschildt P.H., Allard F., Ferguson J., Baron E., Alexander D.R.,
1999,  ApJ, 525, 871

\bibitem{} 
Horne K., Wade R., Szkody, 1986, MNRAS, 219, 791

\bibitem{} 
Israelian G., Rebolo R., Basri G., Casares J., Martin E.L.,  1999,
Nat, 401, 142

\bibitem{}
Kurucz R.L., 1979, ApJS, 40, 1

\bibitem{}
Linnell A.P., Hubeny I., 1994, ApJ, 434, 738

\bibitem{}
Lucy L.B., 1967, Z. Astrophysik, 65, 89

\bibitem{}
Linnell A.P., Hubeny I., 1996, ApJ, 471, 958

\bibitem{} 
Marsh T.R., Robinson E.L., Wood J.H., 19994, MNRAS, 266, 137

\bibitem{} 
Orosz J.A., Bailyn C.D., 1997, ApJ, 477, 876

\bibitem{} 
Orosz J.A., Hauschildt P.H., 2000, A\&A, 364, 265

\bibitem{} 
Orosz J.A., Kuulkers E., van der Klis M., McClintock J.E., Garcia M.R.,  
Callanan P.J., Bailyn C.D., Jain R.K., Remillard R.A.,
2001, ApJ, 555, 489

\bibitem{} 
Orosz J.A., Groot, P.J., van der Klis, M., McClintock J E.,
Garcia M R., Zhao P., Jain R.K., Baily C.D., Remillard R.A.,
2002, ApJ, 568, 845
  

\bibitem{} 
Press W.H., Teukolsky S.A., Vettering V.T., Flannery B.P., 1992, in 
Numerical Recipes in Fortran, 2nd edition, Cambridge University Press, 
Cambridge

\bibitem{} 
Renka R.J., 1988, Algorithm 661: QSHEP3D: Quadratic Shepard method  for  
trivariate interpolation of scattered data, ACM Trans. Math Software, 14, 151

\bibitem{} 
Royer F., Gerbaldi M., Faraggiana R., Gómez A.E.,
2002, A\&A, 381, 105

\bibitem{} 
Shahbaz T., 1998, MNRAS, 298, 153

\bibitem{} 
Shahbaz T., van der Hooft F., Casares J., Charles P.A., van Paradijs J.,  
1999, MNRAS, 306, 89 (S99)

\bibitem{} 
Shepard D., 1968, A two dimensional interpolation for irregularly
spaced data, Proc. 23rd Nat. Conf. ACM Brandon/Systems Press Inc, 
Princeton, 517

\bibitem{} 
Tonry J., Davis M., 1979, AJ, 84, 1511

\bibitem{} 
Torres M.A.P., Casares J., Martinez-Pais I.G., Charles P.A., 
2002, MNRAS, 334, 233

\bibitem{} 
Tjemkes A.S., van Paradijs J., Zuiderwijk E.L., 1986, A\&A, 154, 77

\bibitem{} 
Von Zeipel H., 1924, MNRAS, 84, 655

\bibitem{} 
Welsh W.F., Horne K., Gomer R., 1995, MNRAS, 275, 649

\end{thebibliography}
\end{document}